\documentclass[10pt,twocolumn,showpacs,showkeys,preprintnumbers,amssymb,aps,superscriptaddress,pre]{revtex4-2}

\usepackage{amsmath, amssymb}
\usepackage{graphicx}
\usepackage{color,array}
\usepackage[colorlinks={true}]{hyperref}
\hypersetup{colorlinks=true,linkcolor=red,citecolor=blue,urlcolor=blue}
\usepackage{xcolor} 
\usepackage{color,array}
\usepackage{bm}
\usepackage{subfigure}
\usepackage{amsmath}
\usepackage{lipsum,appendix}
\usepackage{orcidlink}
\usepackage{pstricks}

\begin{document}


\title{Multipartite Entanglement and Quantum Sensing in a Spin-5/2 Heisenberg Molecular Iron(III) Triangle}

\author{Hamid Arian Zad\orcidlink{0000-0002-1348-1777}}
\email{Corresponding author: hamid.arian.zad@upjs.sk}
\address{Department of Theoretical Physics and Astrophysics, Faculty of Science of P. J. \v{S}af{\'a}rik University, Park Angelinum 9, 040 01 Ko\v{s}ice, Slovak Republic}%

\author{Jozef Stre{\v c}ka\orcidlink{0000-0003-1667-6841}}%
\address{Department of Theoretical Physics and Astrophysics, Faculty of Science of P. J. \v{S}af{\'a}rik University, Park Angelinum 9, 040 01 Ko\v{s}ice, Slovak Republic}%

\author{ Winfried Plass\orcidlink{0000-0003-3473-9682}}%
\address{ Institut f\"{u}r Anorganische und Analytische Chemie, Friedrich-Schiller-Universit\"{a}t Jena, 07743 Jena, Germany}%

\date{\today}

\begin{abstract}
This study provides insights into the static and dynamic quantum properties of the trinuclear high-spin iron(III) molecular complex $[\mathrm{Fe}_3\mathrm{Cl}_3(\mathrm{saltag^\mathrm{Br}})(\mathrm{py})_6]\mathrm{ClO}_4$ to be further abbreviated as Fe$_3$. Using exact diagonalization of a spin-5/2 Heisenberg triangle in a magnetic field, we model the corresponding quantum behavior of the molecular compound Fe$_3$. Our rigorous analysis employs various key metrics to explore a rich quantum behavior of this molecular compound. At sufficiently low temperatures, the bipartite negativity reveals that the pairwise entanglement between any pair of iron(III) magnetic ions of the molecular complex Fe$_3$ can be significantly enhanced by a small magnetic field. This enhancement is followed by unconventional step-like changes characterized by a sequence of plateaus and sudden downturns as the magnetic field further increases. A qualitatively similar behavior is also observed in the genuine tripartite entanglement among all three iron(III) magnetic ions in the trinuclear complex Fe$_3$. Notably, the bipartite and tripartite entanglement persist in the molecular complex Fe$_3$ up to moderate temperatures of approximately 30~K and 70~K, respectively. Additionally, we demonstrate the achievement of quantum-enhanced sensitivity by initializing the molecular complex Fe$_3$ in Dicke states. Finally, we investigated a quantum-sensing protocol by applying a local magnetic field specifically to one iron(III) magnetic ion of the molecular compound Fe$_3$ and performing readout sequentially on one of two remaining iron(III) magnetic ions. 
\end{abstract}

\maketitle

\section{Introduction} \label{sec:Introduction}

The field of molecular magnetism \cite{Kahn,Georges2001,Schnack2004,Schnack2005} is rapidly advancing with combining insights from quantum information science, condensed matter physics and chemical engineering \cite{Coronado2020,Pineda2021,Leuenberger2001,Chiesa2024}. Molecular magnets, with their unique features, have promising applications in ultra-high-density storage devices \cite{Qin2009,Baltic2016}, spintronics \cite{Sanvito2011,Fursina2023}, quantum computing \cite{Fursina2023,Song2014,Kandala2017,Carretta2021,Ollitrault2021,Giansiracusa2021,Chizzini2022}, and quantum algorithms \cite{Godfrin2017}. Phenomena like quantum tunneling of magnetization and molecular cooling in the quantum magnets are essential for quantum computing and achieving low temperatures \cite{Thomas1996,Gatteschi2003b,Wu2021,Fix2018}. Molecular magnets with nonlocal correlations also open doors to applications in quantum information technology such as entanglement \cite{Horodecki2009,Bazhanov2018,Vargova2023,Bode2023}, teleportation \cite{Bennett1993,Krauter2013}, and quantum state transferring \cite{Lockyer2022,Ferrando2016,Soria2016}. This growing field provides diverse opportunities that harness the distinctive characteristics of molecular magnets, bridging various scientific disciplines.

Quantum sensing, a key application of quantum technologies \cite{Degen2017,Braun2018,Kugelgen2021}, has witnessed advancements in various physical setups, ranging from smart materials \cite{Chugh2023,Liu2021} and photonic devices \cite{Pirandola2018,Taylor2013} to ion traps \cite{  Baumgart2016,Bohnet2016,Reiter2017,Gilmore2021} and superconducting qubits for the purpose of quantum error correction \cite{Reed2012,Corcoles2015,Danilin2021,Rojkov2022}. The exploration of quantum-enhanced sensitivity is crucial in both critical and Floquet many-body quantum sensors \cite{Mishra2021,Mishra2022,Sarkar2022,Huang2024}. In these systems, the interaction between particles plays a central role, imparting robustness against imperfections. The study of quantum sensors aims to investigate whether projective measurements and their subsequent wave function collapse can be utilized to achieve quantum-enhanced sensitivity, opening new avenues in the field of quantum metrology \cite{Burgarth2015,Smerzi2018,Montenegro2022,Montenegro2023,Montenegro2024,Mihailescu2023}.

Molecular magnets serve as the building blocks for a new era of customizable designer quantum sensors \cite{Troiani2019,Yu2021}, offering tunable functionality and adaptability to various environments \cite{Ghirri2014}. These unique capabilities pave the way for unprecedented levels of sensitivity and spatial resolution, making them useful for a broad spectrum of sensing tasks. From the study of quantum dynamic processes to the precise detection of emergent phenomena, molecular-based quantum materials that exhibit rich magnetic properties present an opportunity to control their spin states using external fields, signaling new era for future quantum devices \cite{Takahata2005,MolecularMagnets,Luis2019,Candini2010}. Additionally, the simultaneous presence of toroidal moments in geometrically frustrated transition metal complexes with $\mathrm{C}_3$ symmetry presents even wider prospects, especially in the presence of strong antiferromagnetic exchange interactions \cite{Crabtree2018,Rao2020,Pister2022}. However, the practical realization of these systems as molecular quantum tools demands a deep comprehension of their electronic structure, magnetism, and their dependencies to factors such as antisymmetric contributions to magnetic exchange (Dzyalozhinsky-Moriya interaction) \cite{Robert2019,Mrozek2021} or single-ion anisotropy \cite{Rinehart2011,Georgiev2022}.

In a recent study, a trinuclear iron(III) molecular complex $[\mathrm{Fe}_3\mathrm{Cl}_3(\mathrm{saltag}^\mathrm{Br})(\mathrm{py})_6]\mathrm{ClO}_4 \\
\{\mathrm{H}_5\mathrm{saltag}^\mathrm{Br} = 1,2,3-\mathrm{tris}[(5-\mathrm{bromo}-\mathrm{salicylidene})\mathrm{amino}]\mathrm{guanidine}\}$ (abbreviated as Fe$_3$) with an antiferromagnetically coupled triangular structure of $\mathrm{C}_3$ symmetry was synthesized and characterized in detail \cite{Plass2023}. Both experimental and theoretical investigations were carried out to explore magnetic properties of this complex \cite{Plass2023}. Motivated by rich magnetic properties of the molecular compound Fe$_3$ and rich quantum properties expected from it, in this work we consider this complex and investigate the genuine bipartite and tripartite entanglement through the behavior of negativity, a concept introduced by Peres and Horodecki \cite{Horodecki2009,Peres1996,Horodecki1996}. This analysis aims to provide insights into the intricate quantum correlations present in the molecular complex Fe$_3$, offering a deeper understanding of its quantum properties.

In this work, we examine the phenomenon of quantum entanglement in the context of the trinuclear high-spin iron(III) complex. 
{ We particularly discuss the theoretical understanding of tripartite entanglement in the molecular compound Fe$_3$ focusing on its stepwise changes along magnetic-field-driven phase transitions. Besides, we also demonstrate quantum-enhanced sensitivity when the molecular complex Fe$_3$ is initialized in Dicke state. A quantum-sensing protocol involves applying a local magnetic field specifically to one of the iron(III) magnetic ions in the Fe$_3$ molecular compound and performing sequential readout on one of the two remaining iron(III) magnetic ions.}

{The structure of the paper is as follows. In Sec. \ref{Sec:Model} we introduce the model and explore the fundamental concepts that underpin quantum sensors based on the molecular magnet Fe$_3$. Then, we investigate thermal bipartite and tripartite negativities of the system with respect to temperature and magnetic field in Sec. \ref{Sec:Negativity}.
In Sec. \ref{Sec:sensing},  the spin-5/2 Heisenberg triangle is firstly initialized in each of the four simplest Dicke states to assess its performance. Subsequently, the quantum spin dynamics are analyzed under the influence of a transverse magnetic field applied to a single subsystem of the model. Next, we investigate the quantum sensing abilities of the molecular complex Fe$_3$ through sequential measurements. Finally, sensitivity is evaluated using the classical Fisher information (CFI) as the quantifier. Section \ref{Sec:Conclusions} contains a brief summary and future outlooks.}

\section{Spin-5/2 Heisenberg triangular cluster}\label{Sec:Model}

Let us examine quantum and magnetic properties of the spin-5/2 Heisenberg triangular cluster, which represents a suitable model system for the trinuclear high-spin iron(III) molecular complex $[\mathrm{Fe}_3\mathrm{Cl}_3(\mathrm{saltag}^\mathrm{Br}) (\mathrm{py})_6]\mathrm{ClO}_4$ referred to as Fe$_3$. It has been convincingly evidenced in Ref. \cite{Plass2023} that the molecular compound Fe$_3$ indeed affords, due to a high-spin state of iron(III) magnetic ions and $\mathrm{C}_3$ symmetry of the underlying superexchange pathways (see Fig. \ref{fig:Fe3_model}), an experimental realization of the spin-5/2 Heisenberg equilateral triangle given by the Hamiltonian:
\begin{figure}[tbp]
	\centering
	\includegraphics[scale=0.15,trim=50 00 10 00, clip]{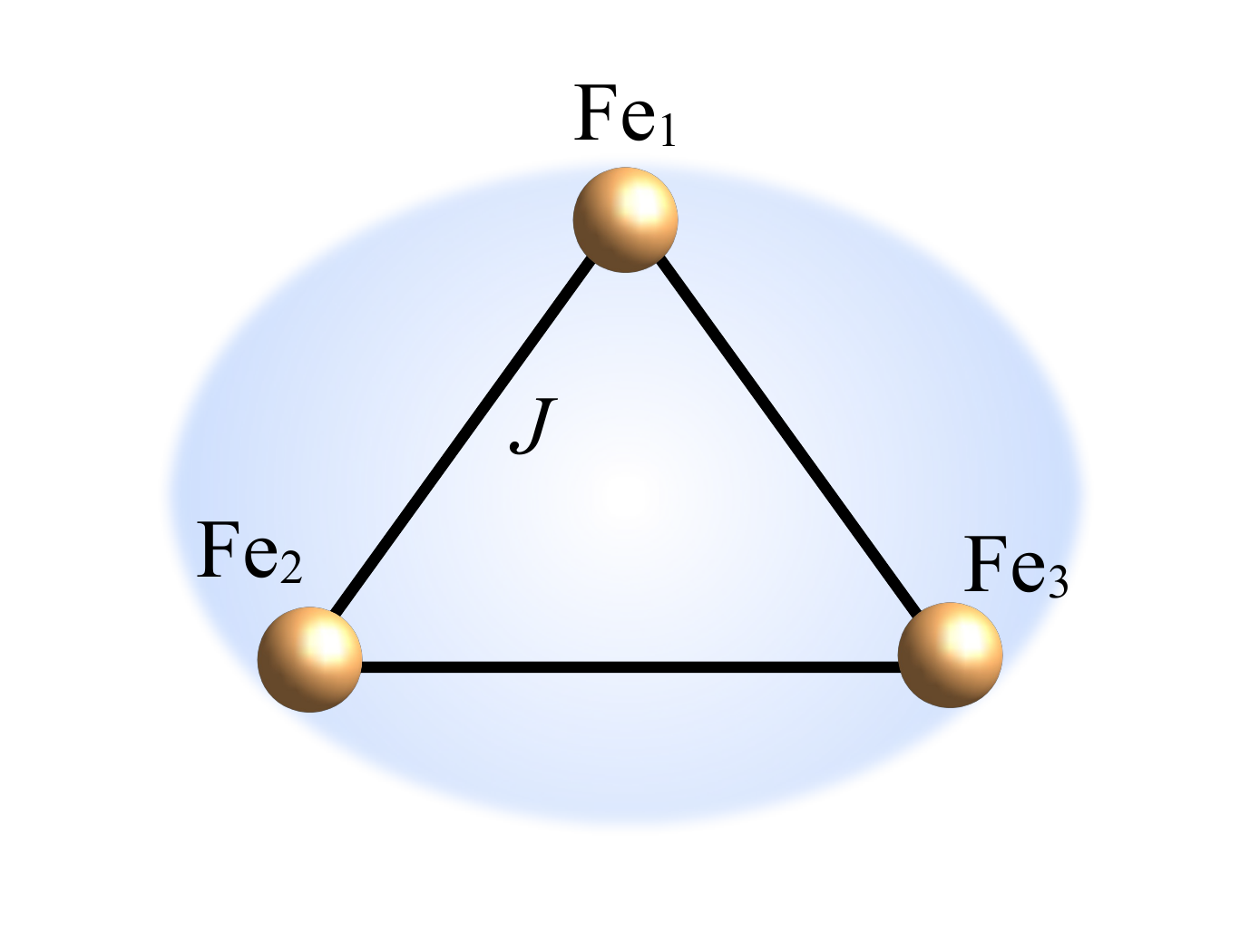} %
	\includegraphics[width=40mm]{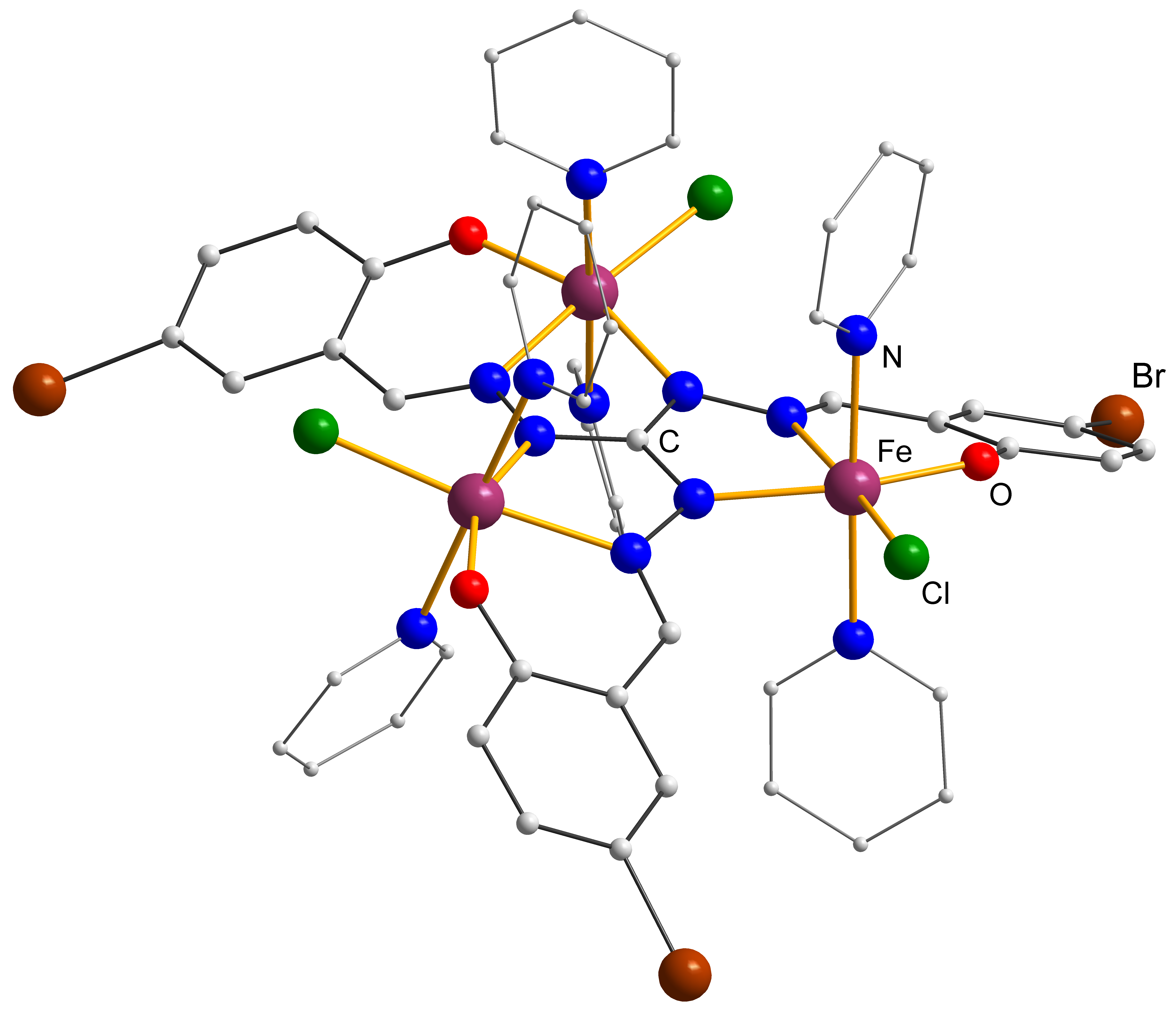} %
	\vspace{-0.25cm}
	\caption{Left panel displays a scheme of three exchange interactions within the trinuclear high-spin iron(III) molecular complex $[\mathrm{Fe}_3\mathrm{Cl}_3(\mathrm{saltag}^\mathrm{Br}) (\mathrm{py})_6] \mathrm{ClO}_4$. Right panel shows the molecular structure of the same complex visualized according to crystallographic data reported in Ref. \cite{Plass2023}.}
	\label{fig:Fe3_model}
\end{figure}
\begin{eqnarray}\label{Eq:Hamiltonian}
	\hat{H} &=& J\left(\hat{\boldsymbol S}_1 \cdot \hat{\boldsymbol S}_2 + \hat{\boldsymbol S}_2 \cdot \hat{\boldsymbol S}_3 + \hat{\boldsymbol S}_1 \cdot \hat{\boldsymbol S}_3\right)  \nonumber \\
	&-& g\mu_B B_z\left(\hat{S}_1^z + \hat{S}_2^z + \hat{S}_3^z\right).
	\label{H}
\end{eqnarray}
Here, $\hat{S}^{\alpha}_j$ ($\alpha =x,y,z$) are conventional spatial components of spin-\(\frac{5}{2}\) operators assigned to three iron(III) metal centers $j=1-3$ of the molecular complex Fe$_3$, the coupling constant $J$ determines a size of intramolecular exchange interactions between the pairs of spin-5/2 iron(III) magnetic ions, $g = 2.0$ is the gyromagnetic ratio of the fully isotropic spin-5/2 iron(III) magnetic ions with zero orbital angular momentum, $\mu_B$ is Bohr magneton and $B_{z}$ is the magnetic field applied along the $z$-direction. According to experimental results and DFT calculations using broken-symmetry states (BS-DFT) \cite{Plass2023}, this coupling constant is antiferromagnetic with a moderate value of $J = 12.56 \, \mathrm{cm}^{-1}$.

By introducing the total spin operator \( \hat{S}_\mathrm{T} = \hat{S}_{1} + \hat{S}_{2}  + \hat{S}_{3}\) and its $z$-component \(\hat{S}_\mathrm{T}^z = \hat{S}_{1}^z + \hat{S}_{2}^z + \hat{S}_{3}^z \), the Hamiltonian (\ref{Eq:Hamiltonian}) of the spin-5/2 Heisenberg triangle can be alternatively expressed within the Kambe projection method \cite{Kambe1950,Sinn1970} in the following form:
\begin{eqnarray}\label{Eq:HamKambe}
	\hat{H} = \frac{J}{2} \left[\hat{S}_\mathrm{T}^2 - (\hat{S}_{1}^2 + \hat{S}_{2}^2 + \hat{S}_{3}^2)\right] -  g\mu_B B_z \hat{S}_\mathrm{T}^{z}.
\end{eqnarray}
Since all spin-5/2 magnetic ions are identical we have $\hat{S}_{1}^2 = \hat{S}_{2}^2 = \hat{S}_{3}^2 = S(S+1) = 35/4$ and $\hat{S}_\mathrm{T}^2 = {S}_\mathrm{T}({S}_\mathrm{T} + 1)$, where the total quantum spin number acquires the following quantized values ${S}_\mathrm{T} = 1/2, 3/2, 5/2, \cdots, 15/2$. The eigenenergies of the Hamiltonian (\ref{Eq:HamKambe}) can be accordingly rewritten in terms of the total quantum spin number \({S}_\mathrm{T}\) and its $z$-component \({S}_\mathrm{T}^{z}\):

\begin{eqnarray}\label{Eq:EigEnKambe}
	E_{S_T, S_T^z} = \frac{J}{2} \left[{S}_\mathrm{T}({S}_\mathrm{T} + 1) - \frac{105}{4}\right] - g\mu_B B_z S_\mathrm{T}^{z}.
\end{eqnarray}
This alternative formula provides a full set of energy eigenvalues of the spin-5/2 Heisenberg triangle  and facilitates the analysis of its magnetic properties within the Kambe coupling scheme \cite{Kambe1950,Sinn1970}.
The rising magnetic field generally causes crossing of energy levels from different sectors of  $z$-component of the total spin \({S}_\mathrm{T}^{z}\), whereby a level-crossing field  for a magnetic-field-driven transition between the sectors with \({S}_\mathrm{T}^{z}\) and \({S}_\mathrm{T}^{z}+1\) is given by the condition \(g\mu_\mathrm{B}B^{z}_{{S}_\mathrm{T}^{z}, {S}_\mathrm{T}^{z}+1}\ = (S_\mathrm{T}+1)J\).
The precise structure of these ground states and their degeneracies are given in \ref{app:appendix1}. 
\begin{figure}[tbp]
	\centering 
	\includegraphics[scale=0.35,trim=00 00 00 00, clip]{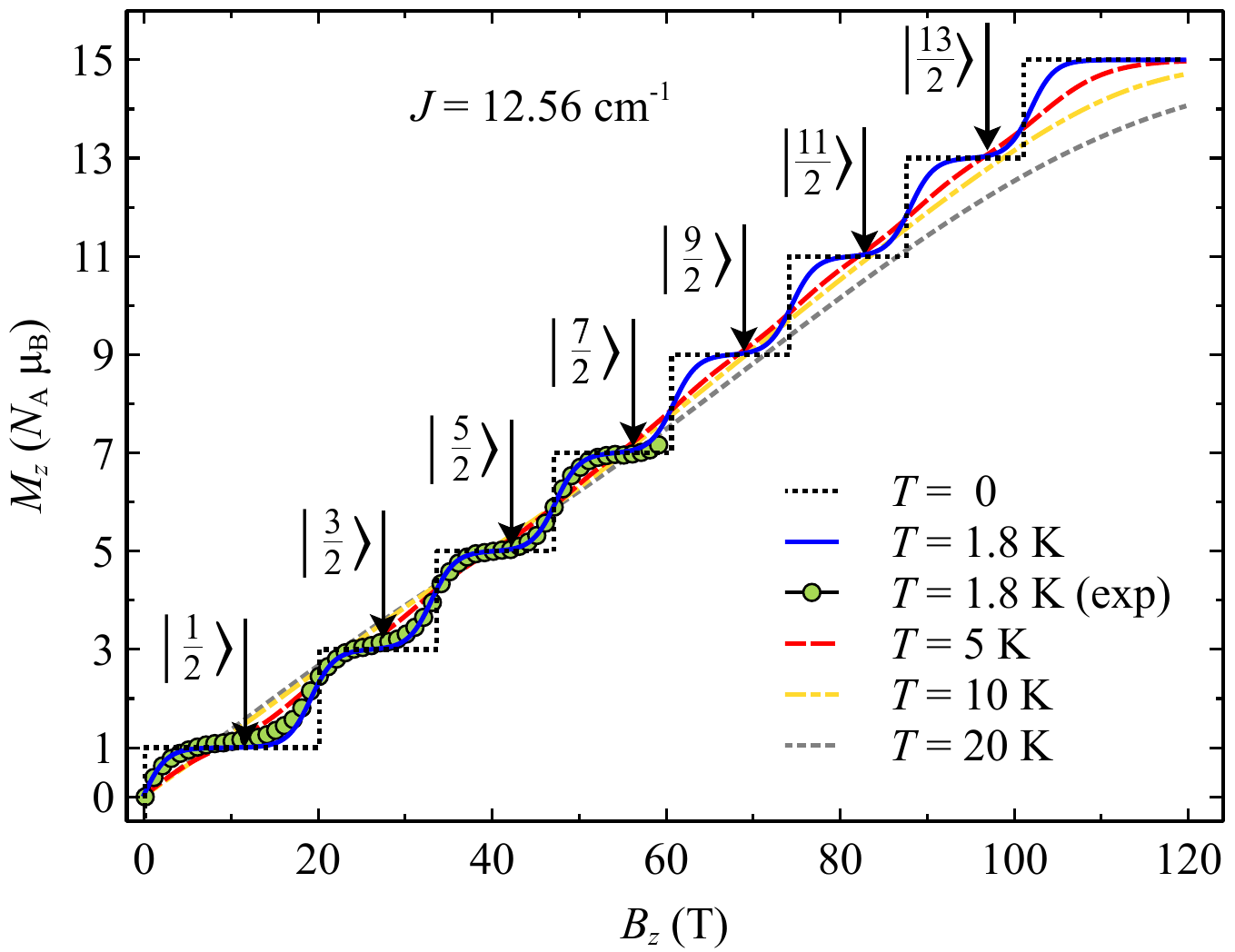} %
	\vspace{-0.25cm}
	\caption{	
		The isothermal magnetization curves of the spin-5/2 Heisenberg triangle calculated for a few selected temperatures by assuming the coupling constant $J = 12.56\, \mathrm{cm}^{-1}$ and the gyromagnetic ratio $g = 2.0$. High-field magnetization data as reported in Ref. \cite{Plass2023} for the molecular complex Fe$_3$ at sufficiently low temperature $T = 1.8\, \mathrm{K}$ are depicted by green circles. The ground-state eigenvectors $|S_\mathrm{T}\rangle$ corresponding to intermediate magnetization plateaus are indicated by arrows.}
	\label{fig:Mag_BT}
\end{figure}

The partition function \( Z \) of the system can be then expressed in terms of the sum of the eigenenergies (\( E_i \)) of the Hamiltonian (\ref{Eq:HamKambe}) as \( Z = \sum_i \exp(-\beta E_i)\), where \( \beta = \frac{1}{k_B T} \), \( k_B \) is the Boltzmann constant and \( T \) is temperature. The Gibbs free energy can be derived from the partition function as \( G = -k_B T \ln Z \) and  one can ultimately obtain the magnetization of the system using the formula \( M_z = - (\partial G / \partial B_z)_T\).

Figure \ref{fig:Mag_BT} displays isothermal magnetization curves of the spin-5/2 Heisenberg triangle for a few selected  temperatures including \( T = 1.8\, \mathrm{K} \), which corresponds to temperature of the pulsed-field magnetization measurement performed on the polycrystalline sample of the molecular complex Fe$_3$ reported in Ref. \cite{Plass2023}. A zero-temperature magnetization curve shown in Fig. \ref{fig:Mag_BT} by a dotted line implies presence of a sequence of intermediate magnetization plateaus at 1/15, 1/5, 1/3, 7/15, 3/5, 11/15, and 13/15 of the saturation magnetization before the magnetization reaches its saturation value. Each intermediate magnetization plateau reflects a quantum state with a given value of the $z$-component of the total spin ${S}_\mathrm{T}^z = 1/2, 3/2, 5/2, \cdots, 15/2$ and the corresponding eigenenergy (\ref{Eq:EigEnKambe}), which becomes energetically most favorable within a specific interval of the magnetic field. The magnetization data, which were experimentally recorded for the molecular compound Fe$_3$ at sufficiently low temperature $1.8$~K up to 60~T shown by green circles, are in a perfect agreement with the respective theoretical prediction depicted by a blue solid line. A comparison between the theoretical and experimental results thus verifies actual existence of first four quantum ground states reflected as the respective fractional magnetization plateaus at 1/15, 1/5, 1/3, and 7/15 of the saturation magnetization. As temperature increases, the intermediate magnetization  plateaus  gradually smear out and the magnetization curve becomes smoother. 

\section{Quantum entanglement}
\label{Sec:Negativity}
The bipartite negativity \cite{Vidal2002} represents one of the most commonly used measures of the pairwise entanglement in the quantum information theory, which  quantifies the amount of nonclassical correlations between two subsystems of a composite quantum system. Owing to the $\mathrm{C}_3$ symmetry of the trinuclear complex Fe$_{3}$, all three reduced density matrices of the spin-5/2 Heisenberg triangle obtained after tracing out degrees of freedom of the third spin ${\rho}_{ij} = \mbox{Tr}_k \rho$ are completely identical ${\rho}_{12} = {\rho}_{23} = {\rho}_{13} \equiv {\rho}_\mathrm{Bip}$. Consequently, all three associated bipartite negativities ${N}_{12} = {N}_{23} = {N}_{13}  \equiv {N}_\mathrm{Bip}$ will become identical as well. To quantify a strength of the bipartite entanglement between any pair of spin-5/2 magnetic ions iron(III) of the trinuclear complex Fe$_{3}$ one can therefore utilize the formula put forward by Vidal and Werner \cite{Vidal2002}, which is defined according to the Peres--Horodecki separability criterion \cite{Peres1996,Horodecki1996} as a sum of absolute values of negative eigenvalues of a partially transposed reduced density matrix $\rho_\mathrm{Bip}^\mathrm{T}$
\begin{equation}\label{Eq:biNeg}
	N_\mathrm{Bip} = \sum_{\lambda_i < 0} |\lambda_i| = \frac{1}{2} \left(\sum_{i=1}^{36} |\lambda_i| - \lambda_i \right).
\end{equation}
Here, \(\lambda_i\) denote the eigenvalues of the reduced density matrix \(\rho_\mathrm{Bip}^{\mathrm{T}}\) that is partially transposed with respect to one of two remaining spins. The reduced density matrix $\rho_\mathrm{Bip}$ and its transpose $\rho_\mathrm{Bip}^\mathrm{T}$ along with the eigenvalues $\lambda_i$ are obtained using exact numerical methods implemented in the {\it QuTip} package \cite{Johansson2012}.

\begin{figure*}[tbp]
	\centering
	\includegraphics[scale=0.35,trim=00 00 00 00, clip]{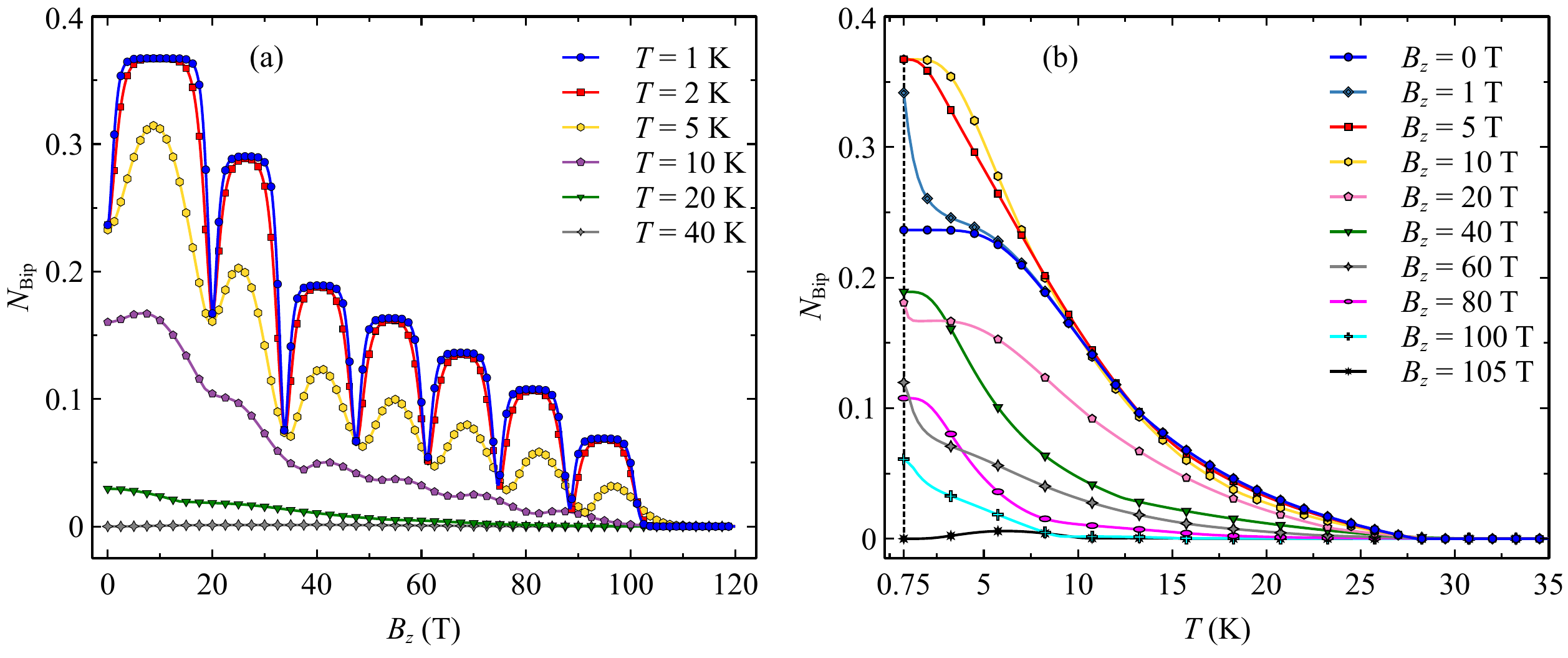} %
	\vspace{-0.5cm}
	\caption{(a) Magnetic-field dependencies of the bipartite negativity ${N}_\mathrm{Bip}$ for a few different values of temperature; (b) temperature dependencies of the bipartite negativity ${N}_\mathrm{Bip}$ for a few selected values of the magnetic field. In both panels we present the theoretical prediction of the bipartite negativity ${N}_\mathrm{Bip}$ for any pair of high-spin iron(III) magnetic ions within the molecular complex Fe$_{3}$ based on the spin-5/2 Heisenberg triangle, which is given by the Hamiltonian (\ref{H}) with the coupling constant $J = 12.56 \, \mathrm{cm}^{-1}$ and the gyromagnetic factor $g = 2.0$ as determined from the previous experiments \cite{Plass2023}.}
	\label{fig:Bip_LN_BT}
\end{figure*}

Figure \ref{fig:Bip_LN_BT} illustrates typical magnetic-field and temperature dependencies of the bipartite negativity ${N}_\mathrm{Bip}$, which measures a strength of the pairwise entanglement between any pair of high-spin iron(III) magnetic ions within the molecular complex Fe$_{3}$. Let us begin by a detailed analysis of the magnetic-field dependencies of the bipartite negativity ${N}_\mathrm{Bip}$, which are exemplified in Figure \ref{fig:Bip_LN_BT}(a) to highlight similarity with magnetization jumps observed in a low-temperature magnetization curve due to magnetic-field-driven phase transitions (cf. Fig. \ref{fig:Bip_LN_BT}(a) with Fig. \ref{fig:Mag_BT}). At sufficiently low temperatures, the bipartite negativity ${N}_\mathrm{Bip}$ reveals that the pairwise entanglement between any pair of iron(III) magnetic ions of the molecular complex Fe$_3$ can be significantly enhanced by a small magnetic field. This enhancement is followed by unconventional step-like changes characterized by a sequence of plateaus and sudden downturns as the magnetic field further increases. The stepwise changes of the bipartite negativity ${N}_\mathrm{Bip}$ evidently relate to magnetic-field-driven phase transitions between the ground states $|S_\mathrm{T}\rangle$ and $|S_\mathrm{T} + 1\rangle$ emergent at the level-crossing field \(g\mu_\mathrm{B}B^{z}\ = (S_\mathrm{T}+1)J\), at which the bipartite negativity ${N}_\mathrm{Bip}$ reaches a pronounced minimum due to a mixed character of the respective ground state. 

The bipartite negativity ${N}_{\mathrm{Bip}}$ is plotted against temperature in Fig. \ref{fig:Bip_LN_BT}(b) for several values of the magnetic field. At zero magnetic field $B = 0$~T (blue curve with circles), the bipartite negativity gradually decreases from a maximum value ${N}_{\mathrm{Bip}} \approx 0.24$ and eventually reaches zero at a threshold temperature $T \approx 27.5\, \mathrm{K}$. At sufficiently small but nonzero magnetic fields $B \lesssim 20$~T, the bipartite negativity decreases monotonically as the temperature increases, starting from a higher initial value of approximately 
${N}_{\mathrm{Bip}} \approx 0.36$, and eventually vanishes at the same threshold temperature. Note that the most thermally persistent bipartite entanglement is accordingly found at a magnetic field of 
$B \lesssim 10$~T, at which the bipartite negativity retains its highest value ${N}_{\mathrm{Bip}} \approx 0.36$ over the broadest temperature range. At higher magnetic fields $B \gtrsim 20$~T, the bipartite negativity ${N}_{\mathrm{Bip}}$ starts from smaller initial value due to a magnetic-field-induced suppression of the quantum entanglement, but the bipartite entanglement persists up to the same threshold temperature $T \approx 27.5\, \mathrm{K}$. 

To determine whether all three spin-5/2 iron(III) magnetic ions in the trinuclear complex Fe$_{3}$ are involved in multipartite quantum correlations, we further examine the tripartite entanglement within the respective spin-5/2 Heisenberg triangle. The tripartite quantum entanglement shared among all three spins is captured by the quantity genuine tripartite negativity ${N}_\mathrm{Trip}$, which is defined as the geometric mean of the bipartite negativities calculated for all three possible bipartitions of the spin-5/2 Heisenberg triangle \cite{Sabin2008}
\begin{equation}\label{Eq:triNeg}
	{N}_\mathrm{Trip} = \sqrt[\raisebox{0.8ex}{\small $\frac{1}{3}$}]{{N}_\mathrm{1|23}\,{N}_\mathrm{2|13}\,{N}_\mathrm{3|12}} = {N}_\mathrm{1|23}.
\end{equation}
In this formula, the genuine tripartite negativity ${N}_\mathrm{Trip}$ is expressed in terms of the bipartite negativities ${N}_\mathrm{A|BC}$ with $\mathrm{A},\mathrm{B},\mathrm{C} = \{1,2,3\}$, which are obtained by summing the absolute values of the negative eigenvalues of the partially transposed density matrix $\rho_\mathrm{ABC}^{\mathrm{T}_\mathrm{A}}$ with respect to the subsystem A of the tripartite system ABC. It should be stressed that the bipartite negativity ${N}_\mathrm{A|BC}$, which measures a degree of the bipartite entanglement between the subsystem A and the combined subsystem BC, is not necessarily equal to the bipartite negativity ${N}_\mathrm{Bip} = {N}_\mathrm{A|B}$, which measures a degree of the bipartite entanglement between the subsystem A and the subsystem B alone of the tripartite system ABC. 

\begin{figure*}
	\centering
	\includegraphics[scale=0.35,trim=00 00 00 00, clip]{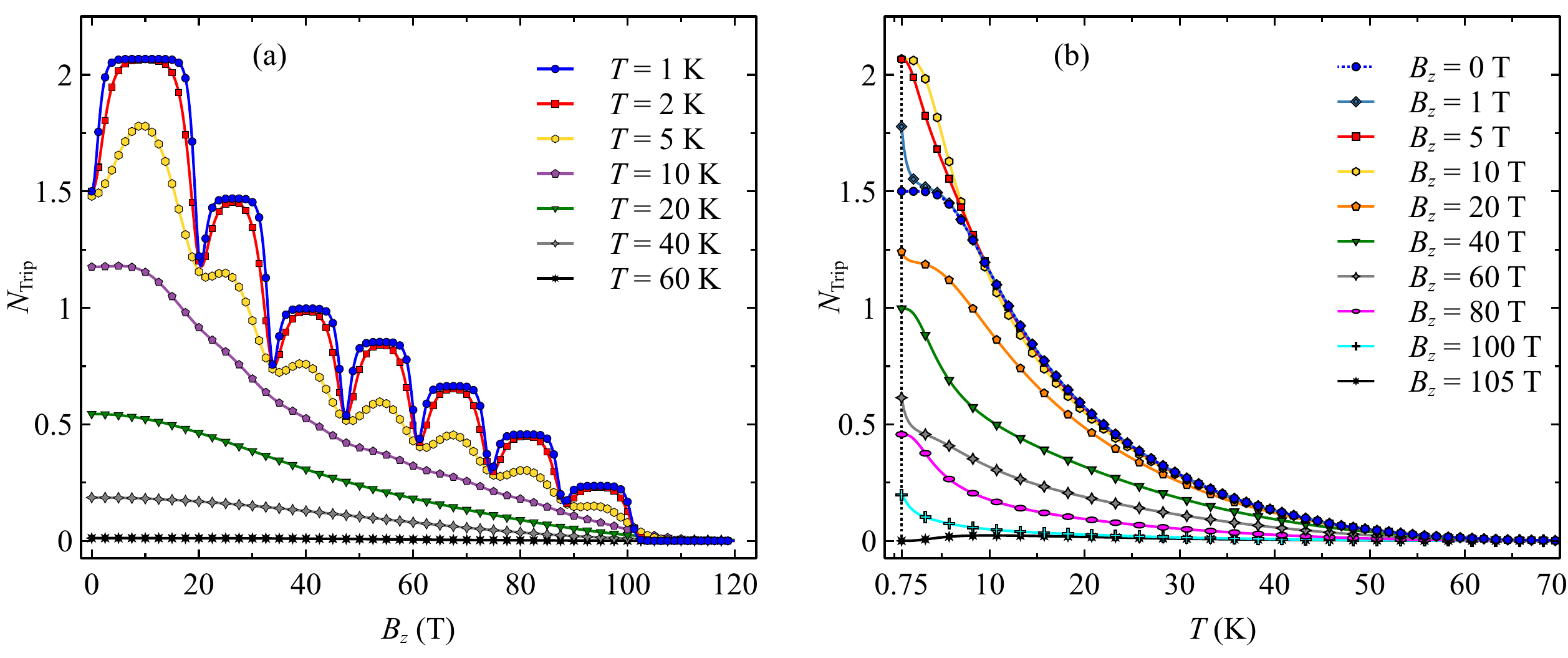} %
	\vspace{-0.5cm}
	\caption{(a) Magnetic-field dependencies of the tripartite negativity ${N}_\mathrm{Trip}$ for a few different values of temperature; (b) temperature dependencies of the tripartite negativity ${N}_\mathrm{Trip}$ for a few selected values of the magnetic field. In both panels we present the theoretical prediction of the tripartite negativity ${N}_\mathrm{Trip}$ for any pair of high-spin iron(III) magnetic ions within the molecular complex Fe$_{3}$ based on the spin-5/2 Heisenberg triangle, which is given by the Hamiltonian (\ref{H}) with the coupling constant $J = 12.56 \, \mathrm{cm}^{-1}$ and the gyromagnetic factor $g = 2.0$ as determined from the previous experiments \cite{Plass2023}.}
	\label{fig:Trip_LN_BT}
\end{figure*}

We now proceed with a theoretical examination of the genuine tripartite negativity ${N}_\mathrm{Trip}$ of the trinuclear complex Fe$_{3}$ as shown in Fig. \ref{fig:Trip_LN_BT}. We note that the {\it QuTip} package \cite{Johansson2012} is used here to numerically obtain the partially transposed density matrix $\rho_\mathrm{ABC}^{\mathrm{T}_\mathrm{A}}$ and its eigenvalues. It is evident from Fig. \ref{fig:Trip_LN_BT}(a) that the behavior of the genuine tripartite negativity closely resembles at low temperatures that of its bipartite counterpart depicted in Fig. \ref{fig:Bip_LN_BT}(a). Specifically, 
the genuine tripartite negativity ${N}_\mathrm{Trip}$  repeatedly exhibits unconventional step-like changes including a series of plateaus and sudden downturns as the magnetic field increases. 
As the temperature rises, this stepwise behavior of the genuine tripartite negativity ${N}_\mathrm{Trip}$ is gradually fading and transitioning into a smooth monotonic decrease with increasing magnetic field (see the lines for $T \gtrsim 10$~K in Fig. \ref{fig:Trip_LN_BT}(a)). Additional insights can be drawn from the temperature dependencies of the genuine tripartite negativity in Fig. \ref{fig:Trip_LN_BT}(b), which reveals that the molecular complex Fe$_{3}$ displays a high degree of tripartite entanglement at low enough temperatures $T \lesssim 10$~K and magnetic fields $B\approx 10\,\mathrm{T}$. Under these conditions, the eigenstate $|1/2 \rangle$ with the lowest possible spin value dominates the overall mixed state. Notably, the tripartite negativity gradually decreases as the temperature increases and eventually vanishes at a significantly higher threshold temperature $T\approx 70\,\mathrm{K}$ compared to the threshold temperature of the bipartite entanglement when exceeding it more than twice. This finding suggests that the trinuclear complex Fe$_{3}$ remains subject to multipartite entanglement even at elevated temperatures and magnetic fields.

\section{Quantum Sensing}\label{Sec:sensing}
To adapt the quantum sensing protocol suggested in Ref. \cite{Montenegro2022} for a real magnetic compound, let us consider the molecular complex Fe$_3$ consisting of three interacting spin-5/2 iron(III) magnetic ions as a probe designed for sensing a local magnetic field $B_x$ acting on one iron(III) magnetic ion via measuring either of the remaining two iron(III) magnetic ions. The Hamiltonian of the spin-5/2 Heisenberg triangle with a local magnetic field acting only on a single spin, which may serve as the probe, can be updated to the following form:
\begin{eqnarray}\label{Eq:H_Bx}
	\hat{{H}^{\prime}} &=& J\left(\hat{\boldsymbol{S}}_1 \cdot \hat{\boldsymbol{S}}_2 + \hat{\boldsymbol{S}}_2 \cdot \hat{\boldsymbol{S}}_3 + \hat{\boldsymbol{S}}_1 \cdot \hat{\boldsymbol{S}}_3\right) 	
	- g\mu_B B_x\hat{S}_1^x.
\end{eqnarray}
In the presence of a nonzero local magnetic field $B_x$, the Dicke states \(|D_{k} \rangle\) serving as an initial state of the probe are not eigenstates of the Hamiltonian $\hat{H}^{\prime}$ and they consequently evolve under the action of the Hamiltonian $\hat{H}^{\prime}$. 

It has been demonstrated in Ref. \cite{Nepomechie} that the Dicke states \( | D_{k} \rangle \) of three qudits can be straightforwardly obtained by applying \( k \) times the total spin-raising operator \( J_+ \) on the basis state with the lowest total spin angular momentum $|-\frac{5}{2},-\frac{5}{2},-\frac{5}{2} \rangle$. More specifically, the Dicke states \( | D_{k} \rangle \) of three spin-5/2 qudits can be defined as follows:
\begin{eqnarray}\label{Eq:dick}
	| D_{k} \rangle = a_{k} \left( J_+ \right)^k \left|-\frac{5}{2},-\frac{5}{2},-\frac{5}{2} \right\rangle,
\end{eqnarray}
where \( k = 0, 1, \ldots, 15 \) and \( a_{k} \) is the normalization factor:
\[
a_{k} = \frac{1}{ k!\sqrt{15 \choose k}}.
\]
The four simplest Dicke states of the spin-5/2 Heisenberg triangle corresponding to $k=0,1,2,3$ read:
\begin{itemize}
	\item[(i)] Fully polarized Dicke state $| D_0 \rangle$ having all spins in the lowest spin state: \\
	\(
	| D_0 \rangle = |-\frac{5}{2}, -\frac{5}{2}, -\frac{5}{2} \rangle;
	\)
	\item[(ii)] Dicke state with one excitation $| D_1 \rangle$ above the fully polarized Dicke state $| D_0 \rangle$: \\
	\(| D_1 \rangle = \frac{1}{\sqrt{3}} (| -\frac{3}{2}, -\frac{5}{2}, -\frac{5}{2} \rangle + |-\frac{5}{2}, -\frac{3}{2}, -\frac{5}{2} \rangle + | -\frac{5}{2}, -\frac{5}{2}, -\frac{3}{2} \rangle);
	\)
	\item[(iii)] Dicke state with two excitations $| D_2 \rangle$ above the fully polarized Dicke state $| D_0 \rangle$:\\
	\(
	|D_2 \rangle = \sqrt{\frac{2}{21}} \big[|-\frac{1}{2}, -\frac{5}{2}, -\frac{5}{2} \rangle + |-\frac{5}{2}, -\frac{1}{2}, -\frac{5}{2} \rangle + |-\frac{5}{2}, -\frac{5}{2}, -\frac{1}{2} \rangle\big] \\
	\qquad\quad + \sqrt{\frac{5}{21}} \big[| -\frac{3}{2}, -\frac{3}{2}, -\frac{5}{2} \rangle + | -\frac{3}{2}, -\frac{5}{2}, -\frac{3}{2} \rangle + |-\frac{5}{2}, -\frac{3}{2}, -\frac{3}{2} \rangle\big];
	\)
	\item[(iv)] Dicke state with three excitations $| D_3 \rangle$ above the fully polarized Dicke state $| D_0 \rangle$: \\
	\(
	| D_3 \rangle = \sqrt{\frac{10}{91}} \big[|-\frac{5}{2}, -\frac{3}{2}, -\frac{1}{2} \rangle + | -\frac{5}{2}, -\frac{1}{2}, -\frac{3}{2} \rangle + | -\frac{3}{2},-\frac{5}{2}, -\frac{1}{2} \rangle  \\
	\qquad\quad + |-\frac{3}{2}, -\frac{1}{2}, -\frac{5}{2} \rangle + | -\frac{1}{2}, -\frac{5}{2}, -\frac{3}{2} \rangle + |-\frac{1}{2}, -\frac{3}{2}, -\frac{5}{2} \rangle\big] \\
	\qquad\quad + \sqrt{\frac{2}{91}} \big[| -\frac{5}{2}, -\frac{5}{2}, \frac{1}{2} \rangle + | -\frac{5}{2}, \frac{1}{2}, -\frac{5}{2} \rangle + | \frac{1}{2}, -\frac{5}{2}, -\frac{5}{2} \rangle\big] + 5\sqrt{\frac{1}{91}} | -\frac{3}{2}, -\frac{3}{2}, -\frac{3}{2} \rangle
	\).
\end{itemize}
The Dicke states generally represent a symmetric quantum superposition of basis states of three spin-5/2 entities with the same value of the total spin angular momentum, whereby the coefficients \( a_{k} \) ensure that these states are normalized and symmetric under any permutation of particles.

To demonstrate how the molecular compound Fe$_3$ can facilitate the implementation of a quantum sensing protocol described in Ref. \cite{Montenegro2022}, we specifically consider a local magnetic field acting on the first spin of the spin-5/2 Heisenberg triangle with the readout performed sequentially on either of the remaining two spins.
\begin{figure*}[tbp]
	\centering
	\includegraphics[scale=0.32,trim=0 00 00 00, clip]{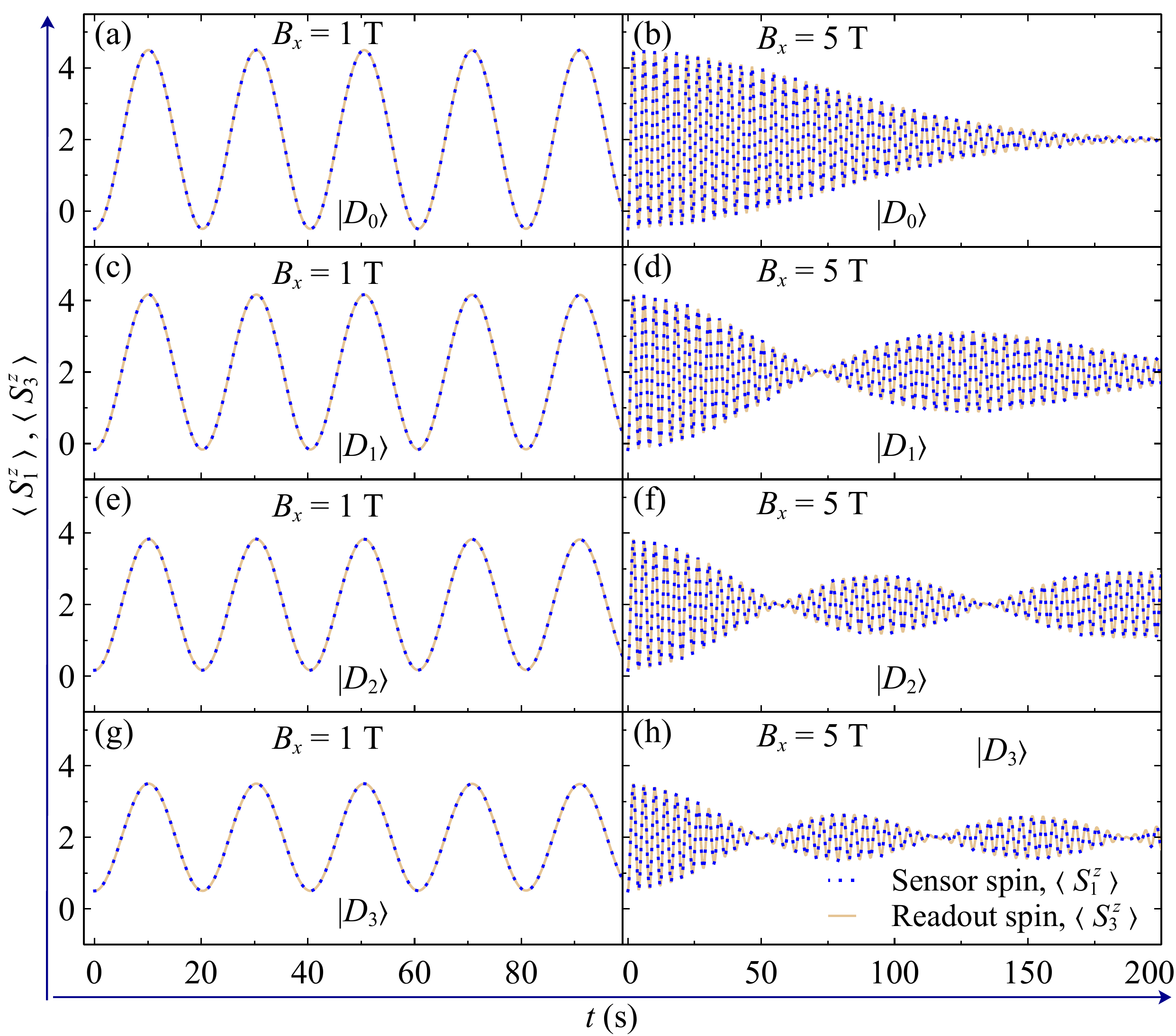} %
	\caption{The time evolution of the local magnetization of the first and third spin of the spin-5/2 Heisenberg triangle with the coupling constant $J = 12.56 \, \mathrm{cm}^{-1}$ and the gyromagnetic factor $g = 2.0$ adjusted for a theoretical modeling of the molecular complex Fe$_3$ \cite{Plass2023} by considering four different initial Dicke states:
		(a)--(b) $| D_0 \rangle$; 
		(c)--(d) $| D_1 \rangle$;
		(e)--(f) $| D_2 \rangle$; 
		(g)--(h) $| D_3 \rangle$.
		In the left panel we suppose a lower magnetic field $B_x = 1 \,\mathrm{T}$, while in the right panel we consider a higher magnetic field $B_x = 5\,\mathrm{T}$. Both spins exhibit almost identical spin dynamics.}
	\label{fig:mz_dynamics}
\end{figure*}
To this end, we first initialize the spin-5/2 Heisenberg triangle in one of the four Dicke states \(| D_0 \rangle \), \(| D_1 \rangle \), \(| D_2 \rangle \) or \(| D_3 \rangle \). The selection of different initial Dicke states allows for the exploration of various dynamical scenarios. When the fully polarized Dicke state \(| D_0 \rangle \) is considered as the initial state, the magnetization evolution of the first and the third spins coincides at sufficiently small magnetic fields $B_x = 1 \,\mathrm{T}$ and 5~T (see Figs. \ref{fig:mz_dynamics}(a), (b)). To bring deeper insight into more challenging conditions for quantum sensing using the molecular compound Fe$_3$, we also initialized the spin-5/2 Heisenberg triangle in the Dicke states \(| D_1 \rangle \), \(| D_2 \rangle \) and \(| D_3 \rangle \) with the form of a symmetric quantum superposition (see Figs. \ref{fig:mz_dynamics}(c)-(h)). It turns out that the coincidence in the spin dynamics of the sensor and readout spin still persists even though the spin dynamics becomes somewhat faster when more complex Dicke states $| D_1 \rangle$, $| D_2 \rangle$, and $| D_3 \rangle$ are used for initialization.

After initializing, the state of the spin-5/2 Heisenberg triangle evolves under the influence of the Hamiltonian \( \hat{H}^{\prime}\) and the dynamics of local magnetization \( \langle S^z_1\rangle \) and \( \langle S^z_3\rangle \) depends on the magnetic field. Fig. \ref{fig:mz_dynamics} illustrates the time evolution of the magnetization dynamics for different initial states under two different magnetic fields $B_x = 1\,\mathrm{T}$ and $B_x = 5\,\mathrm{T}$. Each panel compares the dynamical behavior of the readout spin \( \langle S_3^z \rangle \) (pale lines) and the sensor spin \( \langle S_1^z \rangle \) (blue dotted lines) for two different magnetic fields. Figs. \ref{fig:mz_dynamics}(a) and \ref{fig:mz_dynamics}(b) illustrate the time evolution of magnetization for the initial state \(| D_0 \rangle \) with character of the fully polarized ferromagnetic state. In Fig. \ref{fig:mz_dynamics}(a), the sensor spin shows at a relatively small magnetic field \( B_x = 1 \, \mathrm{T} \) oscillatory harmonic dependence and the readout spin closely follows this dependence. Fig. \ref{fig:mz_dynamics}(b) shows the time evolution of the same quantities at the higher magnetic field \( B_x = 5 \, \mathrm{T} \), for which the readout spin still mimics time evolution of the sensor spin but the oscillations are of higher frequency and less regular with gradually decreasing (and increasing) amplitude over the time. This scenario suggests more complex dynamical behavior as the magnetic field strengthens though the readout spin still continues to follow the sensor spin.

Figs. \ref{fig:mz_dynamics}(c)-\ref{fig:mz_dynamics}(h) depict the time evolution of magnetization for more complex initial Dicke states being subject to quantum superposition of states.
At sufficiently small magnetic fields, the sensor and readout spins exhibit regular synchronized harmonic oscillations in time as demonstrated in Figs. \ref{fig:mz_dynamics}(c), \ref{fig:mz_dynamics}(e) and \ref{fig:mz_dynamics}(g) for the particular case with \( B_x = 1 \, \mathrm{T} \) and a few different Dicke states $| D_1 \rangle$, $| D_2 \rangle$, and $| D_3 \rangle$. Contrary to this, the local magnetization exhibits at the higher magnetic field \( B_x = 5 \, \mathrm{T} \) a more complex oscillatory dynamics with a clear amplitude modulation. Initially, the oscillations have a larger amplitude, but the amplitude decreases in time indicating a form of damped oscillation followed by a partial revival at later times. While the amplitude never reaches its initial value, this revival suggests a dynamic interference or rephasing effect potentially due to an external magnetic field. Although the dynamical oscillations of the sensor and readout spins are much more complex, the sensor and readout spins maintain a synchronized oscillations at small up to moderate magnetic fields across various initial states what highlights potential of the molecular compound Fe$_3$ for indirect measurement of the magnetic field \( B_x \) through the readout spin dynamics.

\subsection{Sequential measurement}
The sequential measurement protocol originally demonstrated in Ref. \cite{Montenegro2022} offers an innovative approach to optimize sensing efficiency by exploiting measurement-induced dynamics. According to the protocol's description given in Ref. \cite{Montenegro2022}, let us outline the key steps of this protocol in the context of single-parameter estimation when specifically focusing on estimating a local trasnverse magnetic field \(B_x\) in absence of the longitudinal magnetic field \(B_z = 0\). First, the system is initialized in one of four lowest Dicke states \(|D_0 \rangle\), \(| D_1 \rangle\), \(| D_2 \rangle\), or \(| D_3 \rangle\).  Then, the system undergoes the following sequential measurements:
\begin{enumerate}
	\item {Free Evolution:} The system evolves freely for a duration of time \(\tau_i\) according to the Hamiltonian \(\hat{H^{\prime}}\). Mathematically, this evolution can be represented as:
	\[
	|\psi_{\tau_i}^{(i)} \rangle = e^{-i\tau_i \hat{H^{\prime}}} |\psi^{(i)}(0) \rangle
	\]
	where \(|\psi^{(i)}(0) \rangle = |D_k\rangle\) is the initial state of the molecular compound Fe$_3$ before the time evolution within a given time interval $\tau_i$ takes place.
	\item {Measurement Outcome:}
	\begin{itemize}
		\item[(a)] After the free evolution, a measurement of the local magnetization is performed on the third spin of the molecular complex \(\mathrm{Fe}_3\).
		\item[(b)] The measurement outcome \(\gamma_i\), which corresponds to one of six available spin states \(| m_z \rangle\) with $m_z \in \{\pm \frac{1}{2}, \pm \frac{3}{2}, \pm \frac{5}{2}\}$, 
		occurs with the probability \(p_{\gamma_i}^{(i)}\).
		\item[(c)] The probabilities \(p_{\gamma_i}^{(i)}\) are calculated as
		\(
		p_{\gamma_i}^{(i)} = \langle \psi_{\tau_i}^{(i)} | \Pi^{\gamma}_3 | \psi_{\tau_i}^{(i)} \rangle
		\),
		where \(\Pi^{\gamma}_3\) is the projection operator corresponding to the measurement outcome \(| m_z \rangle\) on the third spin, i.e., \(\Pi^{m_z}_3 = \mathbf{I}_1 \otimes \mathbf{I}_2 \otimes (|m_z\rangle\langle m_z|)_3  \) ($\mathbf{I}$ stands for the identity operator).
	\end{itemize}
	\item {State Update:} After obtaining the outcome \(\gamma_i\) the state vector reduces to:
	\( | \psi^{(i+1)}(0) \rangle = \frac{1}{\sqrt{p_{\gamma_i}^{(i)}}} \Pi^{\gamma}_3 | \psi^{(i)}(\tau_i) \rangle
	\), where \(\Pi^{\gamma}_3\) is the projection operator corresponding to the measurement outcome.
	
	\item {Iteration:} The new state from the previous step becomes the initial state for the next iteration. Steps (ii) and (iii) are repeated for a total of \(n_{\mathrm{seq}}\) measurements, each separated by intervals of free evolution time $\tau_i$.
	
	\item{Data Collection:}
	A trajectory of length \(n_{\mathrm{seq}}\) of outcomes \({\boldsymbol \gamma} = (\gamma_1, \gamma_2, \ldots, \gamma_{n_{\mathrm{seq}}})\) is gathered.
	
	\item{Probe Reset and Repetition:}
	After gathering a trajectory, the probe is reset to initial state, and the entire process is repeated to generate a new trajectory.
\end{enumerate}

This protocol efficiently utilizes time resources by avoiding the overhead time required for reinitialization after each measurement. Instead, a sequence of measurements is performed successively on the same probe with intermittent-free evolution periods. This strategy allows for effective data gathering while minimizing initialization overhead. Further details of this protocol can be found in the original work \cite{Montenegro2022} and its subsequent studies \cite{Montenegro2023, Montenegro2024}.

\subsection{Parameter estimation}
Let us now proceed with calculating the classical Fisher information (CFI) \( F(B_x) \) \cite{Yuan2020}, which will be used for estimating a local magnetic field \( B_x \) using the measurement basis of the third spin as described by the conventional precision bound. The CFI is a concept from classical statistics that quantifies the amount of information that an observable random variable provides about an unknown parameter. It measures the sensitivity of the likelihood function to changes in the parameter. In our case, the CFI is given by \cite{Montenegro2022}:
\begin{equation}\label{Eq:CFI}
	F(B_x) = \sum_{\boldsymbol{\gamma}} \left[ \frac{1}{P_{\boldsymbol{\gamma}}} \left( \frac{dP_{\boldsymbol{\gamma}}}{dB_x} \right)^2 \right],
\end{equation}
where \( P_{\boldsymbol{\gamma}}=\Pi_{i=1}^{n_\mathrm{seq}}p_{\gamma_i}^{(i)} \) represents the probability of obtaining a particular trajectory \( \boldsymbol{\gamma} \). The summation $\sum_{\boldsymbol{\gamma}}$ runs over all \( 2^{n_{\mathrm{seq}}} \) configurations from the initial states to the final state \(| \frac{5}{2}, \frac{5}{2}, \frac{5}{2} \rangle \).

\begin{figure}[tbp]
	\centering
	\includegraphics[scale=0.245,trim=10 00 00 00, clip]{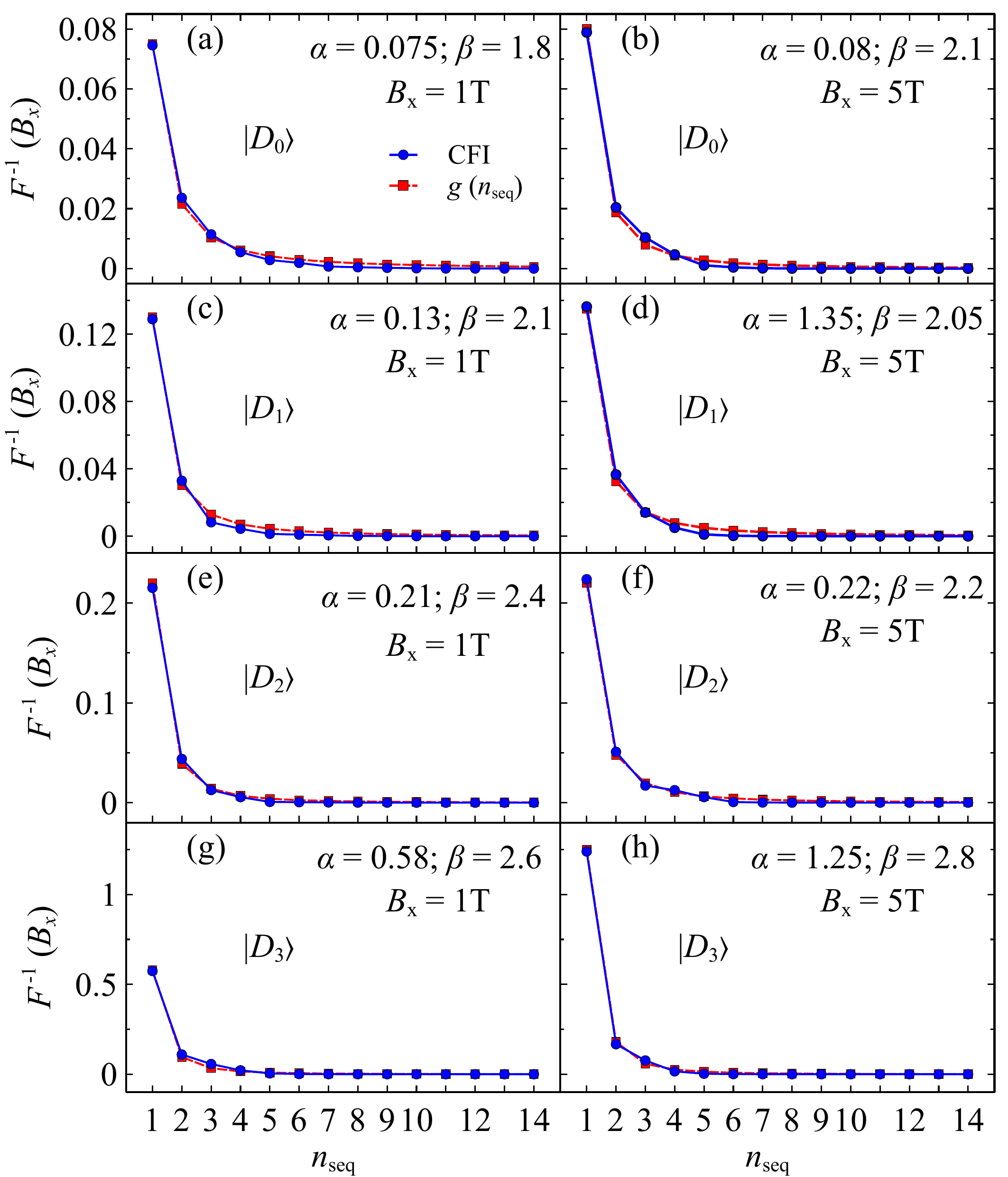} %
	\vspace{-0.5cm}
	\caption{The inverse of classical Fisher information $F^{-1}$ versus the number of sequential measurements $n_{\mathrm{seq}}$ performed at the third spin as calculated according to Eq. (\ref{Eq:CFI}) for the spin-5/2 Heisenberg triangle with the coupling constant $J = 12.56 \, \mathrm{cm}^{-1}$ and the gyromagnetic factor $g = 2.0$. This parameter set was adjusted for a theoretical modeling of the molecular complex Fe$_3$ \cite{Plass2023} initialized in one of four Dicke states: (a), (b) $| D_0 \rangle$; (c), (d) $| D_1 \rangle$; (e), (f) $| D_2 \rangle$; (g), (h) $| D_3 \rangle$. The left and right panels show the respective results for $F^{-1}$ at two selected values of the magnetic field $B_x = 1 \mathrm{T}$ and $B_x = 5 \mathrm{T}$, respectively. The fitting functions $g(n_{\mathrm{seq}})=\alpha n_{\mathrm{seq}}^{-\beta}$ with the exponent coefficient $\beta > 1$ are also shown.}
	\label{fig:CFI}
\end{figure}

To observe the effect of sequential measurements on sensing precision of the molecular compound Fe$_3$, we present in Fig. \ref{fig:CFI} the inverse of the CFI \( F^{-1} \) as a function of the number of sequential measurements \(n_{\mathrm{seq}}\) performed at the third spin. The panels compare results for different initial states and two selected magnetic-field strengths. For example, Fig. \ref{fig:CFI}(a) and (b) shows \( F^{-1} \) for the initial ferromagnetic Dicke state \(| D_0 \rangle\) by considering the local magnetic field \( B_x = 1 \mathrm{T} \). and \( B_x = 5 \mathrm{T} \), respectively. Evidently, the inverse of CFI in Figs. \ref{fig:CFI}(a) and \ref{fig:CFI}(b) decreases rapidly with increasing the number of sequential measurements \( n_{\mathrm{seq}} \). This observation would suggest that the precision of sensing improves significantly when performing greater number of sequential measurements. The fitting function \( g(n_{\mathrm{seq}}) = \alpha n_{\mathrm{seq}}^{-\beta} + \epsilon \) with \( \epsilon \) being negligible and \( \beta \) consistently greater than 1, suggests that the sequential measurements enable a quantum-enhanced sensitivity beyond the classical limit \cite{Montenegro2022}.

Figs. \ref{fig:CFI}(c)-\ref{fig:CFI}(h) display the inverse of CFI for more complex initial Dicke states \(| D_1 \rangle\), \(| D_2 \rangle\), or \(| D_3 \rangle\) with character of a quantum superposition of states by considering two different local magnetic fields \( B_x = 1 \mathrm{T} \) (left panel) and \( B_x = 5 \mathrm{T} \) (right panel). It is evident from Figs. \ref{fig:CFI}(c)-\ref{fig:CFI}(h) that the inverse of CFI decreases significantly as the number of sequential measurements \( n_\mathrm{seq} \) increases. This behavior supports the conclusion that sequential measurements using the trinuclear molecular complex Fe$_3$ substantially enhance sensing precision regardless of the complexity of the initial Dicke state with the nature of quantum superposition of states. The decay of inverse CFI with increasing the number of sequential measurements can be satisfactorily fitted with a power-law dependence with the exponent \( \beta > 1 \), which in turn confirms the advantage of sequential measurements for the high-precision sensing of the magnetic field. 

\section{Conclusions}\label{Sec:Conclusions}

In this article, we have thoroughly investigated the quantum features of the trinuclear molecular complex Fe$_3$, which represents an exemplary experimental realization of a spin-5/2 Heisenberg triangle with a moderately strong antiferromagnetic coupling constant $J = 12.56 \, \mathrm{cm}^{-1}$ and an isotropic gyromagnetic factor $g = 2.0$. The low-temperature magnetization curve of the molecular compound Fe$_3$ is expected to exhibit a sequence of fractional magnetization plateaus at 1/15, 1/5, 1/3, 7/15, 3/5, 11/15, and 13/15 of the saturation magnetization. Among these, the first four intermediate plateaus were experimentally confirmed through high-field magnetization measurements up to 60~T. 

Our findings demonstrate that the molecular complex Fe$_3$ exhibits sizable bipartite and tripartite  entanglement, which are thermally quite robust and persist up to temperatures around 30 K and 70 K, respectively. In addition, we have found that a remarkable synchronization is detected between the readout and sensor spin when a local magnetic field is applied to the first spin of the molecular complex Fe$_3$ and the local magnetization of the third spin is measured. This synchronization between the spin dynamics of the readout and sensor spin suggests that the measurement of the third spin can provide remote information about the local field at the first spin, which makes the molecular complex Fe$_3$ an effective magnetic sensor for quantum information processing.

Besides, our results indicate that the molecular complex Fe$_3$ has the potential for achieving quantum-enhanced sensitivity beyond the classical limit. In fact, the significant impact of sequential measurements on the high-precision sensing of local magnetic field has been clearly demonstrated. By plotting the inverse of CFI against the number of sequential measurements for various initial Dicke states and magnetic fields, we have furnished a rigorous proof for consistent and rather steep decrease in the inverse value of CFI with increasing number of sequential measurements. This trend highlights the substantial enhancement in sensing precision achieved through sequential measurements on the molecular complex Fe$_3$.

The outcomes of this work could inspire further research on other molecular magnets, more specifically, in the area of quantized magnetization plateaus, entanglement, and quantum sensing in various transition-metal complexes. This comprehensive exploration provides valuable insights into several intriguing quantum phenomena exhibited by the molecular complex Fe$_3$ and highlights the potential of related molecular quantum magnets for near-term applications in quantum technologies.

\section*{Acknowledgments}
H. Arian Zad acknowledges the financial support provided under the postdoctoral fellowship program of P. J. \v{S}af\'{a}rik University in Ko\v{s}ice, Slovakia. This research was funded by Slovak Research and Development Agency under the contract No. APVV-20-0150 and  The Ministry of Education, Research, Development and Youth of the Slovak Republic under the grant number VEGA 1/0695/23.

 \noindent
 \onecolumngrid
\appendix
\section{}\label{app:appendix1}
\noindent
The lowest-energy eigenstates of the spin-5/2 Heisenberg triangle given by the Hamiltonian (\ref{Eq:Hamiltonian}) within all possible spin multiplets. The eigenenergies $E_{S_T} = E_{S_T, S_T^z = S_T}$ of the lowest-energy eigenstates are unambiguously given by Eq. (\ref{Eq:EigEnKambe}) and the corresponding eigenvector $|S_T\rangle$ are classified according to the total quantum spin number $S_T$ (the upper indices are used if the relevant eigenstate is degenerate):

\noindent
\begin{eqnarray*}
	E_{15/2} &=& \frac{75}{4}J - g\mu_\mathrm{B}\frac{15}{2}B,
	\\
	|15/2\rangle &=& \left| 5/2, 5/2, 5/2 \right\rangle,
    \\
	E_{13/2} &=& \frac{45}{4}J - g\mu_\mathrm{B}\frac{13}{2}B, 
    \\
	|13/2\rangle^\mathrm{i} &=& -0.58 \left| 5/2, 5/2, 3/2 \right\rangle + 0.71 \left| 5/2, 3/2, 5/2 \right\rangle, \\
	|13/2\rangle^\mathrm{ii} &=& -0.58 \left| 5/2, 5/2, 3/2 \right\rangle + 0.71 \left| 3/2, 5/2, 5/2 \right\rangle,
     \\
	E_{11/2} &=& \frac{19}{4}J - g\mu_\mathrm{B}\frac{11}{2}B, 
    \\
	|11/2\rangle^\mathrm{i} &=& 0.38 \left| 5/2, 5/2, 1/2 \right\rangle  -0.49 \left| 3/2, 5/2, 3/2 \right\rangle + 0.58 \left| 1/2, 5/2, 5/2 \right\rangle, \\
	|11/2\rangle^\mathrm{ii} &=& 0.59 \left| 5/2, 5/2, 1/2 \right\rangle -0.42 \left| 5/2, 3/2, 3/2 \right\rangle -0.39 \left| 3/2, 5/2, 3/2 \right\rangle \\
	&& + 0.47 \left| 3/2, 3/2, 5/2 \right\rangle, \\
	|11/2\rangle^\mathrm{iii} &=& 0.38 \left| 5/2, 5/2, 1/2 \right\rangle -0.53 \left| 5/2, 3/2, 3/2 \right\rangle  + 0.47 \left| 5/2, 1/2, 5/2 \right\rangle, 
    \\
	E_{9/2} &=& -\frac{3}{4}J - g\mu_\mathrm{B}\frac{9}{2}B, 
    \\
	|9/2\rangle^\mathrm{i} &=& -0.23 \left| 5/2, 5/2, -1/2 \right\rangle + 0.28 \left| 3/2, 5/2, 1/2 \right\rangle \\
	&& -0.35 \left| 1/2, 5/2, 3/2 \right\rangle + 0.50 \left| -1/2, 5/2, 5/2 \right\rangle,
	\\
	|9/2\rangle^\mathrm{ii} &=& -0.52 \left| 5/2, 5/2, -1/2 \right\rangle + 0.27 \left| 5/2, 3/2, 1/2 \right\rangle \\
	&& + 0.42 \left| 3/2, 5/2, 1/2 \right\rangle -0.29 \left| 3/2, 3/2, 3/2 \right\rangle \\
	&& -0.26 \left| 1/2, 5/2, 3/2 \right\rangle + 0.35 \left| 1/2, 3/2, 5/2 \right\rangle, \\
	        \end{eqnarray*}
	\begin{eqnarray*}
	|9/2\rangle^\mathrm{iii} &=& -0.52 \left| 5/2, 5/2, -1/2 \right\rangle + 0.54 \left| 5/2, 3/2, 1/2 \right\rangle \\
	&& -0.26 \left| 5/2, 1/2, 3/2 \right\rangle + 0.21 \left| 3/2, 5/2, 1/2 \right\rangle \\
	&& -0.29 \left| 3/2, 3/2, 3/2 \right\rangle + 0.25 \left| 3/2, 1/2, 5/2 \right\rangle, \\
	|9/2\rangle^\mathrm{iv} &=& -0.23 \left| 5/2, 5/2, -1/2 \right\rangle + 0.36 \left| 5/2, 3/2, 1/2 \right\rangle \\
	&& -0.35 \left| 5/2, 1/2, 3/2 \right\rangle + 0.28 \left| 5/2, -1/2, 5/2 \right\rangle, 
        \\
	E_{7/2} &=& -\frac{21}{4}J - g\mu_\mathrm{B}\frac{7}{2}B, 
    \\
	|7/2\rangle^\mathrm{i} &=& 0.13 | 5/2, 5/2, -3/2 \rangle - 0.13 | 3/2, 5/2, -1/2 \rangle + 0.14 | 1/2, 5/2, 1/2 \rangle \\
	&& - 0.21 | -1/2, 5/2, 3/2 \rangle + 0.45 | -3/2, 5/2, 5/2 \rangle, \\
	|7/2\rangle^\mathrm{ii} &=& 0.42 | 5/2, 5/2, -3/2 \rangle - 0.16 | 5/2, 3/2, -1/2 \rangle - 0.32 | 3/2, 5/2, -1/2 \rangle \\
	&& + 0.14 | 3/2, 3/2, 1/2 \rangle + 0.22 | 1/2, 5/2, 1/2 \rangle - 0.16 | 1/2, 3/2, 3/2 \rangle \\
	&& - 0.17 | -1/2, 5/2, 3/2 \rangle + 0.27 | -1/2, 3/2, 5/2 \rangle, \\
	|7/2\rangle^\mathrm{iii} &=& 0.59 | 5/2, 5/2, -3/2 \rangle - 0.44 | 5/2, 3/2, -1/2 \rangle + 0.16 | 5/2, 1/2, 1/2 \rangle \\
	&& - 0.30 | 3/2, 5/2, -1/2 \rangle + 0.27 | 3/2, 3/2, 1/2 \rangle - 0.13 | 3/2, 1/2, 3/2 \rangle \\
	&& + 0.10 | 1/2, 5/2, 1/2 \rangle - 0.15 | 1/2, 3/2, 3/2 \rangle + 0.14 | 1/2, 1/2, 5/2 \rangle, \\
	|7/2\rangle^\mathrm{iv} &=& 0.42 | 5/2, 5/2, -3/2 \rangle - 0.47 | 5/2, 3/2, -1/2 \rangle + 0.34 | 5/2, 1/2, 1/2 \rangle \\
	&& - 0.13 | 5/2, -1/2, 3/2 \rangle - 0.11 | 3/2, 5/2, -1/2 \rangle + 0.14 | 3/2, 3/2, 1/2 \rangle \\
	&& - 0.14 | 3/2, 1/2, 3/2 \rangle + 0.11 | 3/2, -1/2, 5/2 \rangle, \\
	|7/2\rangle^\mathrm{v} &=& 0.13 | 5/2, 5/2, -3/2 \rangle - 0.20 | 5/2, 3/2, -1/2 \rangle + 0.22 | 5/2, 1/2, 1/2 \rangle \\
	&& - 0.17 | 5/2, -1/2, 3/2 \rangle + 0.17 | 5/2, -3/2, 5/2 \rangle, 
	\\
	E_{5/2} &=& -\frac{35}{4}J - g\mu_\mathrm{B}\frac{5}{2}B,
	 \\
	|5/2\rangle^\mathrm{i} &=& -0.06 | 5/2, 5/2, -5/2 \rangle + 0.05 | 3/2, 5/2, -3/2 \rangle - 0.03 | 1/2, 5/2, -1/2 \rangle \\
	&& + 0.04 | -1/2, 5/2, 1/2 \rangle - 0.10 | -3/2, 5/2, 3/2 \rangle + 0.41 | -5/2, 5/2, 5/2 \rangle, \\
	|5/2\rangle^\mathrm{ii} &=& -0.29 | 5/2, 5/2, -5/2 \rangle + 0.07 | 5/2, 3/2, -3/2 \rangle + 0.18 | 3/2, 5/2, -3/2 \rangle \\
	&& -0.05 | 3/2, 3/2, -1/2 \rangle - 0.10 | 1/2, 5/2, -1/2 \rangle + 0.04 | 1/2, 3/2, 1/2 \rangle \\
	&& + 0.08 | -1/2, 5/2, 1/2 \rangle - 0.07 | -1/2, 3/2, 3/2 \rangle - 0.10 | -3/2, 5/2, 3/2 \rangle \\
	&& + 0.18 | -3/2, 3/2, 5/2 \rangle, 
	\\
	|5/2\rangle^\mathrm{iii} &=& -0.59 | 5/2, 5/2, -5/2 \rangle + 0.29 | 5/2, 3/2, -3/2 \rangle - 0.08 | 5/2, 1/2, -1/2 \rangle \\
	&& + 0.27 | 3/2, 5/2, -3/2 \rangle - 0.14 | 3/2, 3/2, -1/2 \rangle + 0.05 | 3/2, 1/2, 1/2 \rangle \\
	&& - 0.10 | 1/2, 5/2, -1/2 \rangle + 0.09 | 1/2, 3/2, 1/2 \rangle - 0.05 | 1/2, 1/2, 3/2 \rangle \\
	&& + 0.04 | -1/2, 5/2, 1/2 \rangle - 0.07 | -1/2, 3/2, 3/2 \rangle + 0.07 | -1/2, 1/2, 5/2 \rangle, \\
	|5/2\rangle^\mathrm{iv} &=& -0.59 | 5/2, 5/2, -5/2 \rangle + 0.44 | 5/2, 3/2, -3/2 \rangle - 0.24 | 5/2, 1/2, -1/2 \rangle \\
	&& + 0.06 | 5/2, -1/2, 1/2 \rangle + 0.18 | 3/2, 5/2, -3/2 \rangle - 0.14 | 3/2, 3/2, -1/2 \rangle \\
	&& + 0.10 | 3/2, 1/2, 1/2 \rangle - 0.04 | 3/2, -1/2, 3/2 \rangle - 0.03 | 1/2, 5/2, -1/2 \rangle \\
	&& + 0.04 | 1/2, 3/2, 1/2 \rangle - 0.05 | 1/2, 1/2, 3/2 \rangle + 0.04 | 1/2, -1/2, 5/2 \rangle, \\
	|5/2\rangle^\mathrm{v} &=& -0.29 | 5/2, 5/2, -5/2 \rangle + 0.29 | 5/2, 3/2, -3/2 \rangle - 0.24 | 5/2, 1/2, -1/2 \rangle \\
	&& + 0.13 | 5/2, -1/2, 1/2 \rangle - 0.10 | 5/2, -3/2, 3/2 \rangle + 0.07 | 5/2, -5/2, 5/2 \rangle, \\
	|5/2\rangle^\mathrm{vi} &=&  -0.06|5/2, 5/2, -5/2\rangle + 0.07|5/2, 3/2, -3/2\rangle + -0.08|5/2, 1/2, -1/2\rangle \\
	&& + 0.06|5/2, -1/2, 1/2\rangle + -0.05|5/2, -3/2, 3/2\rangle + 0.10|5/2, -5/2, 5/2\rangle,
	 \\
	E_{3/2} &=& -\frac{45}{4}J - g\mu_\mathrm{B}\frac{3}{2}B,
	\\
	|3/2 \rangle^\mathrm{i} &=&  0.20 |5/2, 3/2, -5/2\rangle - 0.25 |5/2, 1/2, -3/2\rangle + 0.28 |5/2, -1/2, -1/2\rangle \\
	&& - 0.32 |5/2, -3/2, 1/2\rangle + 0.22 |5/2, -5/2, 3/2\rangle - 0.16 |3/2, 5/2, -5/2\rangle \\
	&& + 0.11 |3/2, 3/2, -3/2\rangle - 0.04 |3/2, 1/2, -1/2\rangle - 0.04 |3/2, -1/2, 1/2\rangle \\
	&& + 0.08 |3/2, -3/2, 3/2\rangle - 0.16 |3/2, -5/2, 5/2\rangle + 0.04 |1/2, 5/2, -3/2\rangle \\
	&& - 0.06 |1/2, 3/2, -1/2\rangle + 0.06 |1/2, 1/2, 1/2\rangle - 0.06 |1/2, -1/2, 3/2\rangle \\
	&& + 0.04 |1/2, -3/2, 5/2\rangle, \\
		\end{eqnarray*}
	\begin{eqnarray*}
	| 3/2 \rangle^\mathrm{ii} &=&  0.26 |5/2, 3/2, -5/2\rangle - 0.20 |5/2, 1/2, -3/2\rangle + 0.15 |5/2, -1/2, -1/2\rangle \\
	&& - 0.13 |5/2, -3/2, 1/2\rangle + 0.07 |5/2, -5/2, 3/2\rangle - 0.20 |3/2, 5/2, -5/2\rangle \\
	&& + 0.06 |3/2, 1/2, -1/2\rangle - 0.06 |3/2, -1/2, 1/2\rangle + 0.04 |3/2, -3/2, 3/2\rangle \\
	&& + 0.12 |1/2, 5/2, -3/2\rangle - 0.05 |1/2, 3/2, -1/2\rangle - 0.10 |-1/2, 5/2, -1/2\rangle \\
	&& + 0.06 |-1/2, 3/2, 1/2\rangle + 0.17 |-3/2, 5/2, 1/2\rangle - 0.12 |-3/2, 3/2, 3/2\rangle \\
	&& - 0.38 |-5/2, 5/2, 3/2\rangle + 0.38 |-5/2, 3/2, 5/2\rangle - 0.05 |3/2, -5/2, 5/2\rangle, \\
	| 3/2\rangle^\mathrm{iii} &=&  0.61 |5/2, 3/2, -5/2\rangle - 0.57 |5/2, 1/2, -3/2\rangle + 0.47 |5/2, -1/2, -1/2\rangle \\
	&& - 0.41 |5/2, -3/2, 1/2\rangle + 0.22 |5/2, -5/2, 3/2\rangle - 0.49 |3/2, 5/2, -5/2\rangle \\
	&& + 0.11 |3/2, 3/2, -3/2\rangle + 0.11 |3/2, 1/2, -1/2\rangle - 0.18 |3/2, -1/2, 1/2\rangle  \\
	&& - 0.16 |3/2, -5/2, 5/2\rangle + 0.22 |1/2, 5/2, -3/2\rangle - 0.17 |1/2, 3/2, -1/2\rangle  \\
	&& - 0.12 |-1/2, 5/2, -1/2\rangle + 0.16 |-1/2, 3/2, 1/2\rangle - 0.08 |-1/2, 1/2, 3/2\rangle \\
	&& + 0.10 |-3/2, 5/2, 1/2\rangle - 0.15 |-3/2, 3/2, 3/2\rangle + 0.16 |-3/2, 1/2, 5/2\rangle \\
	&& + 0.06 |1/2, 1/2, 1/2\rangle + 0.14 |3/2, -3/2, 3/2\rangle, \\
	| 3/2 \rangle^\mathrm{iv} &=&  0.57 |5/2, 3/2, -5/2\rangle - 0.63 |5/2, 1/2, -3/2\rangle + 0.60 |5/2, -1/2, -1/2\rangle \\
	&& - 0.57 |5/2, -3/2, 1/2\rangle + 0.32 |5/2, -5/2, 3/2\rangle - 0.46 |3/2, 5/2, -5/2\rangle \\
	&& + 0.20 |3/2, 3/2, -3/2\rangle + 0.02 |3/2, 1/2, -1/2\rangle - 0.17 |3/2, -1/2, 1/2\rangle  \\
	&& - 0.23 |3/2, -5/2, 5/2\rangle + 0.16 |1/2, 5/2, -3/2\rangle - 0.18 |1/2, 3/2, -1/2\rangle  \\
	&& - 0.06 |1/2, -1/2, 3/2\rangle - 0.04 |-1/2, 5/2, -1/2\rangle + 0.08 |-1/2, 3/2, 1/2\rangle \\
	&& - 0.08 |-1/2, 1/2, 3/2\rangle + 0.07 |-1/2, -1/2, 5/2\rangle  + 0.20 |3/2, -3/2, 3/2\rangle \\
	&& + 0.12 |1/2, 1/2, 1/2\rangle, \\
	| 3/2 \rangle^\mathrm{v} &=&  -0.57 |5/2, 1/2, -5/2\rangle + 0.58 |5/2, -1/2, -3/2\rangle - 0.47 |5/2, -3/2, -1/2\rangle \\
	&& + 0.34 |5/2, -5/2, 1/2\rangle + 0.38 |3/2, 3/2, -5/2\rangle - 0.17 |3/2, 1/2, -3/2\rangle \\
	&& - 0.06 |3/2, -1/2, -1/2\rangle + 0.19 |3/2, -3/2, 1/2\rangle - 0.18 |3/2, -5/2, 3/2\rangle  \\
	&& - 0.09 |1/2, 3/2, -3/2\rangle + 0.15 |1/2, 1/2, -1/2\rangle - 0.10 |1/2, -1/2, 1/2\rangle  \\
	&& + 0.12 |1/2, -5/2, 5/2\rangle + 0.19 |-1/2, 5/2, -3/2\rangle - 0.08 |-1/2, 3/2, -1/2\rangle \\
	&& - 0.06 |-1/2, 1/2, 1/2\rangle + 0.12 |-1/2, -1/2, 3/2\rangle - 0.10 |-1/2, -3/2, 5/2\rangle \\
	&& - 0.25 |-3/2, 5/2, -1/2\rangle + 0.21 |-3/2, 3/2, 1/2\rangle - 0.13 |-3/2, 1/2, 3/2\rangle \\
	&& + 0.29 |-5/2, 5/2, 1/2\rangle - 0.40 |-5/2, 3/2, 3/2\rangle + 0.35 |-5/2, 1/2, 5/2\rangle \\
	&& - 0.20 |1/2, 5/2, -5/2\rangle - 0.01 |1/2, -3/2, 3/2\rangle, \\
	| 3/2 \rangle^\mathrm{vi} &=&  -0.43 |5/2, 1/2, -5/2\rangle + 0.49 |5/2, -1/2, -3/2\rangle - 0.45 |5/2, -3/2, -1/2\rangle \\
	&& + 0.38 |5/2, -5/2, 1/2\rangle + 0.29 |3/2, 3/2, -5/2\rangle - 0.17 |3/2, 1/2, -3/2\rangle \\
	&& + 0.15 |3/2, -3/2, 1/2\rangle - 0.20 |3/2, -5/2, 3/2\rangle - 0.15 |1/2, 5/2, -5/2\rangle  \\
	&& + 0.11 |1/2, 1/2, -1/2\rangle - 0.11 |1/2, -1/2, 1/2\rangle + 0.03 |1/2, -3/2, 3/2\rangle  \\
	&& + 0.12 |-1/2, 5/2, -3/2\rangle - 0.08 |-1/2, 3/2, -1/2\rangle + 0.09 |-1/2, -1/2, 3/2\rangle \\
	&& - 0.15 |-1/2, -3/2, 5/2\rangle - 0.11 |-3/2, 5/2, -1/2\rangle + 0.14 |-3/2, 3/2, 1/2\rangle \\
	&& - 0.17 |-3/2, 1/2, 3/2\rangle + 0.16 |-3/2, -1/2, 5/2\rangle - 0.03 |1/2, 3/2, -3/2\rangle \\
	&& + 0.14 |1/2, -5/2, 5/2\rangle + 0.14 |1/2, -5/2, 5/2\rangle, 
	\\
	E_{1/2} &=& -\frac{51}{4}J - g\mu_\mathrm{B}\frac{1}{2}B, 
    \\
	| 1/2 \rangle^\mathrm{i} &=& -0.57 |5/2, 1/2, -5/2\rangle + 0.58 |5/2, -1/2, -3/2\rangle - 0.47 |5/2, -3/2, -1/2\rangle \\
	&& + 0.34 |5/2, -5/2, 1/2\rangle + 0.38 |3/2, 3/2, -5/2\rangle - 0.17 |3/2, 1/2, -3/2\rangle \\
	&& - 0.06 |3/2, -1/2, -1/2\rangle + 0.19 |3/2, -3/2, 1/2\rangle - 0.18 |3/2, -5/2, 3/2\rangle \\
	&& - 0.20 |1/2, 5/2, -5/2\rangle - 0.09 |1/2, 3/2, -3/2\rangle + 0.15 |1/2, 1/2, -1/2\rangle \\
	&& - 0.10 |1/2, -1/2, 1/2\rangle - 0.01 |1/2, -3/2, 3/2\rangle + 0.12 |1/2, -5/2, 5/2\rangle \\
	&& + 0.19 |-1/2, 5/2, -3/2\rangle - 0.08 |-1/2, 3/2, -1/2\rangle - 0.06 |-1/2, 1/2, 1/2\rangle \\
	&& + 0.12 |-1/2, -1/2, 3/2\rangle - 0.10 |-1/2, -3/2, 5/2\rangle - 0.25 |-3/2, 5/2, -1/2\rangle \\
	&& + 0.21 |-3/2, 3/2, 1/2\rangle - 0.13 |-3/2, 1/2, 3/2\rangle + 0.29 |-5/2, 5/2, 1/2\rangle \\
	&& - 0.40 |-5/2, 3/2, 3/2\rangle + 0.35 |-5/2, 1/2, 5/2\rangle, \\
\end{eqnarray*}
\begin{eqnarray*}
	| 1/2 \rangle^\mathrm{ii} &=& -0.43 |5/2, 1/2, -5/2\rangle + 0.49 |5/2, -1/2, -3/2\rangle - 0.45 |5/2, -3/2, -1/2\rangle \\
	&& + 0.38 |5/2, -5/2, 1/2\rangle + 0.29 |3/2, 3/2, -5/2\rangle - 0.17 |3/2, 1/2, -3/2\rangle \\
	&& + 0.15 |3/2, -3/2, 1/2\rangle - 0.20 |3/2, -5/2, 3/2\rangle - 0.15 |1/2, 5/2, -5/2\rangle \\
	&& - 0.03 |1/2, 3/2, -3/2\rangle + 0.11 |1/2, 1/2, -1/2\rangle - 0.11 |1/2, -1/2, 1/2\rangle \\
	&& + 0.03 |1/2, -3/2, 3/2\rangle + 0.14 |1/2, -5/2, 5/2\rangle + 0.12 |-1/2, 5/2, -3/2\rangle \\
	&& - 0.08 |-1/2, 3/2, -1/2\rangle + 0.09 |-1/2, -1/2, 3/2\rangle - 0.15 |-1/2, -3/2, 5/2\rangle \\
	&& - 0.11 |-3/2, 5/2, -1/2\rangle + 0.14 |-3/2, 3/2, 1/2\rangle - 0.17 |-3/2, 1/2, 3/2\rangle \\
	&& + 0.16 |-3/2, -1/2, 5/2\rangle.
	\\
\end{eqnarray*}
\twocolumngrid

\end{document}